# Fermi surface in the absence of a Fermi liquid in the Kondo insulator SmB$_6$

M. Hartstein[1*], W. H. Toews[2*], Y.-T. Hsu[1*], B. Zeng[3], X. Chen[1],
M. Ciomaga Hatnean[4], Q. R. Zhang[3], S. Nakamura[5], A. S. Padgett[6],
G. Rodway-Gant[1,7], J. Berk[1], M. K. Kingston[1], G. H. Zhang[1,8], M. K. Chan[9],
S. Yamashita[10], T. Sakakibara[5], Y. Takano[6], J.-H. Park[3], L. Balicas[3],
N. Harrison[9], N. Shitsevalova[11], G. Balakrishnan[4], G. G. Lonzarich[1],
R. W. Hill[2], M. Sutherland[1†], & Suchitra E. Sebastian[1†]

[1]Cavendish Laboratory, Cambridge University, Cambridge CB3 0HE, UK,
[2]Department of Physics and Astronomy, University of Waterloo, Waterloo, ON N2L 3G1, Canada,
[3]National High Magnetic Field Laboratory, Tallahassee, FL 32310, USA,
[4]Department of Physics, University of Warwick, Coventry, CV4 7AL, UK,
[5]The Institute for Solid State Physics, The University of Tokyo, Kashiwa, Chiba 277-8581, Japan,
[6]Department of Physics, University of Florida, Gainesville, FL 32611, USA,
[7]Department of Physics, Oxford University, Oxford, OX1 3PU, UK,
[8]Department of Physics, Massachusetts Institute of Technology, Cambridge, MA 02139, USA,
[9]National High Magnetic Field Laboratory, LANL, Los Alamos, NM 87504, USA,
[10]Department of Chemistry, Osaka University, Toyonaka, Osaka 560-0043, Japan,
[11]The National Academy of Sciences of Ukraine, Kiev 03680, Ukraine.

*These authors contributed equally to this work.

**The search for a Fermi surface in the absence of a conventional Fermi liquid has thus far yielded very few potential candidates. Among promising materials are spin-frustrated Mott insulators near the insulator-metal transition, where theory predicts a Fermi surface associated with neutral low energy excitations. Here we reveal another route to experimentally realise a Fermi surface in the absence of a Fermi liquid by the experimental study of a Kondo insulator $SmB_6$ positioned close to the insulator-metal transition. We present experimental signatures down to low temperatures ($\ll 1$ K) associated with a Fermi surface in the bulk, including a sizeable linear specific heat coefficient, and on the application of a finite magnetic field, bulk magnetic quantum oscillations, finite quantum oscillatory entropy, and substantial enhancement in thermal conductivity well below the charge gap energy scale. Thus, the weight of evidence indicates that despite an extreme instance of Fermi liquid breakdown in Kondo insulating $SmB_6$, a Fermi surface arises from novel itinerant low energy excitations that couple to magnetic fields, but not weak DC electric fields.**

The $f$-electron system $SmB_6$, which has been recently proposed to be a topological insulator characterised by a conducting surface [1, 2, 3, 4, 5, 6, 7], has been long known to exhibit Kondo insulating behaviour characterised by a collective $f$-$d$ hybridisation charge gap in the bulk. The bulk charge gap is evidenced in experiments such as infrared absorption, inelastic neutron scattering, optical conductivity, electron tunnelling, intermediate-temperature specific heat capacity, and electrical resistivity [8]. $SmB_6$ is further positioned in the close vicinity of the Kondo insulator transition to a metallic phase, requiring as little as 40 kbar for metallisation [9, 10, 11]. The surprising observation of quantum oscillations in the magnetisation unaccompanied by oscillations in the electrical resistance of $SmB_6$ was reported in ref. [12, 13]. While ref. [12] interpreted these quantum oscillations in the framework of a two-dimensional Fermi surface from a conducting surface layer, ref. [13] in contrast associated them with a three-dimensional Fermi surface from the insulating bulk. Here we test for three-dimensional bulk Fermi surface character associated with the measured quantum oscillations in $SmB_6$, and probe for quantitative correspondence with an itinerant band of in-gap low energy excitations using complementary experimental techniques.

Fig. 1a shows a sample of quantum oscillations in the magnetic torque before any background subtraction, measured in a floating zone-grown crystal of $SmB_6$, revealing large oscillations dominant against the measured background, with prominent high frequency oscillations at high magnetic fields as shown by the inset. The correspondence of the measured quantum oscillations to a three-dimensional ellipsoidal Fermi surface geometry characteristic of metallic hexaborides (Fig. 1c, refs. [13, 14, 15, 16]) is seen from the extended angular dependence of the measured quantum oscillations for tilt angles spanning both the [011]-[001] and [001]-[111]-[110] planes (Fig. 1b). Angular dependent quantum oscillation data is shown for both floating zone-grown and flux-grown single crystals, the observed angular dependence is independent of the orientation of exposed crystal surfaces in both types of samples, in contrast to the expectation for an origin of quantum oscillations from a surface layer. We note that the observation of quantum oscillation frequencies that span the entire angular range is inconsistent with a two-dimensional Fermi surface geometry, for which quantum oscillation frequencies would be expected to vanish at tilt angles corresponding to open Fermi surface orbits (see figures provided in ref. [17]). We next establish the three-dimensional Fermi surface we access to correspond to the bulk volume of the sample by a quantitative inspection of the observed large amplitude of quantum oscillations in the magnetisation (Fig. 1). We choose for comparison the measured lowest frequency quantum oscillations, which are closest to the zero phase (infinite field) limit. Given the small size of the Fermi surface corresponding to the lowest frequency quantum oscillations only occupying $0.1\%$ of the Brillouin zone, the corresponding carrier density is very low. This low carrier density leads to theoretical predictions of a small magnitude of quantum oscil-lation amplitude (in units of $\mu_\text{B}$ per unit cell) in the case of a two-dimensional Fermi surface originating from the surface atomic layer, as well as in the case of a three-dimensional Fermi surface originating from the insulating bulk (Fig. 1g, Methods). We first compare the measured quantum oscillation amplitude per unit cell assuming an origin from the entire bulk with the theoretical prediction. We find good agreement within an order of magnitude of the measured quantum oscillation amplitude (in units of $\mu_\text{B}$ per unit cell) with (i) the theoretical prediction for a three-dimensional bulk Fermi surface made using the Lifshitz-Kosevich theory (Fig. 1e, g, Methods), and (ii) an experimental comparison with the three-dimensional bulk Fermi surface in metallic $LaB_6$ (Fig. 1d, Methods). In contrast, were the observed quantum oscillations to originate solely from the surface atomic layer, these would correspond to a very large amplitude in $\mu_\text{B}$ per surface unit cell, given that surface unit cells constitute only a small fraction $\approx 10^{-6}$ of the total number of unit cells (Fig. 1f). We thus show that were quantum oscillations to originate from a surface atomic layer, the theoretical prediction of the maximum possible amplitude of quantum oscillations per surface unit cell would be several orders of magnitude smaller than the experimentally observed quantum oscillation amplitude per surface unit cell, ruling out such an origin of quantum oscillations reported here. Quantum oscillations in $SmB_6$ are also observed using capacitive Faraday magnetometry measurements in superconducting magnetic fields up to 14 T (see figure in ref. [17]), and bulk magnetic susceptibility measurements in pulsed magnetic fields (see figure in ref. [17]).

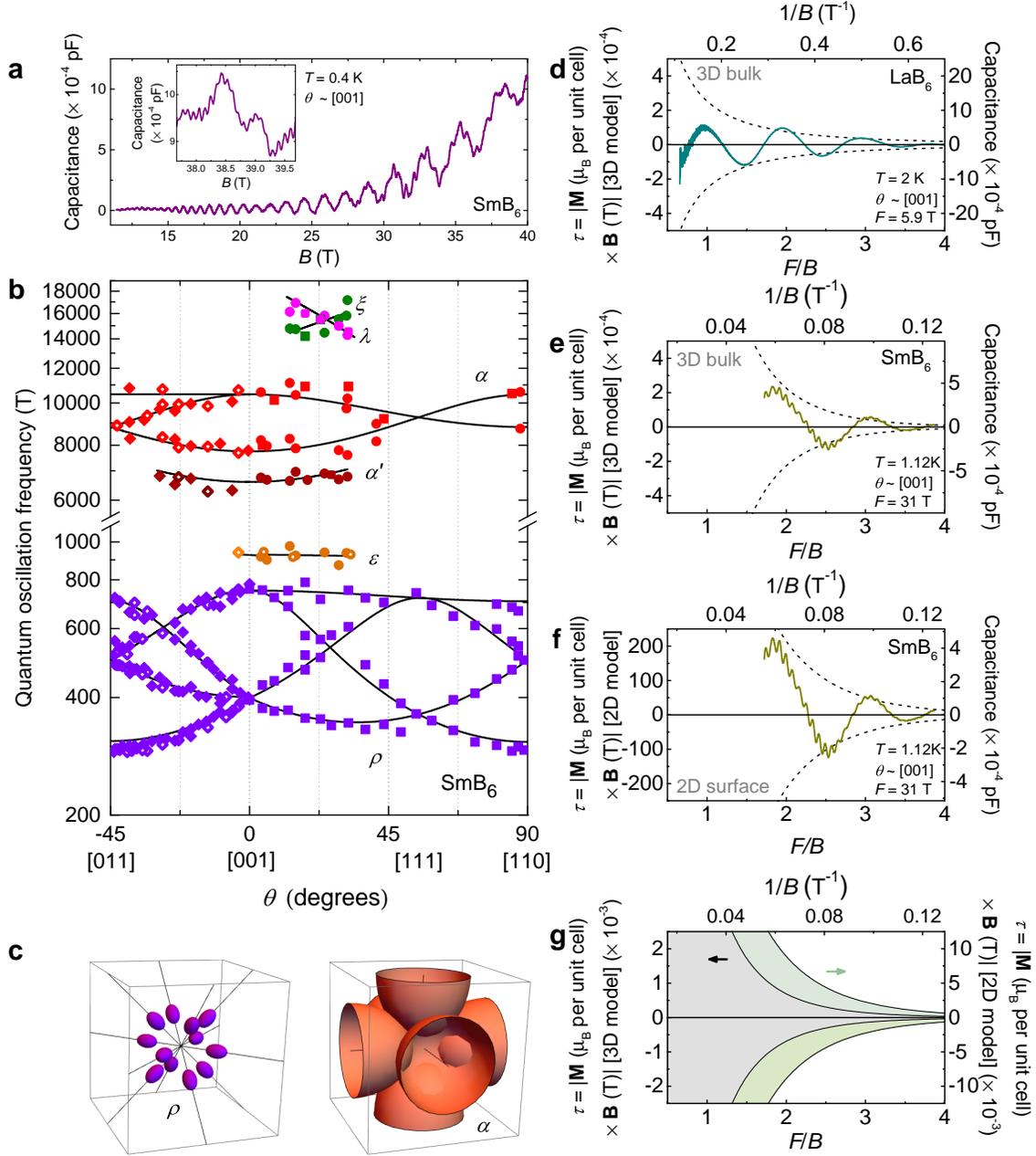

Figure 1: **Comparison of quantum oscillations in SmB$_6$ with three-dimensional bulk Fermi surface model.** **a** shows the measured oscillations in the magnetic torque for a floating zone-grown crystal before any background subtraction with the inset giving a magnified view of the high frequency oscillations visible at high magnetic fields. **b**, Angular dependent quantum oscillation measurements in the [011]-[001] rotation plane in the field range 8 T $< B <$ 35 T on two crystals, and in the [001]-[111]-[110] rotation plane in the field range 11 T $< B <$ 40 T on two other, to complement previous angular dependent measurements reported in ref. [13]. Open and closed circles represent data from floating zone-grown crystals, open and closed diamonds, and squares represent data from flux-grown crystals. Throughout, $B = \mu_0 H$, where $H$ is the applied magnetic field. (Next page.)

Figure 1: (Previous page.) Angular dependence of observed quantum oscillations is in good agreement with the three-dimensional ellipsoidal model characteristic of rare-earth metallic hexaborides and proposed in ref. [13] (shown by fit lines). **c**, Twelve ellipsoidal electron pockets along the <110> directions, corresponding to the fit to the $\rho$ frequencies, and three large ellipsoidal electron pockets along the <100> directions, corresponding to the fit to the $\alpha$ frequencies in **b**. **e**, The quantum oscillatory magnetic moment (in $\mu_B$ per unit cell) corresponding to the measured lowest frequency oscillations in SmB$_6$ (labelled $\rho$' also corresponding to small ellipsoids [13, 14, 15, 16]). **d**, The quantum oscillatory magnetic moment measured using a very similar experimental setup in LaB$_6$ yields a value close to SmB$_6$ assuming a bulk origin in both cases. Dashed lines represent magnetic field dependence of the quantum oscillation amplitude from an exponential damping (Dingle) fit. **f**, Size of the magnetic moment corresponding to the measured lowest frequency oscillations in SmB$_6$ were they to originate from only the surface. **g**, Theoretical Lifshitz-Kosevich estimate for the quantum oscillatory magnetic moment (in $\mu_B$ per unit cell) including the angular anisotropy term, Dingle and spin-splitting damping factors (Methods) indicated for a bulk origin (left-hand axis), and for a surface origin (right-hand axis) of quantum oscillations. Good order of magnitude agreement is seen with experiment assuming a bulk origin (**d**, **e**), whereas the predicted theoretical maximum size is several orders of magnitude smaller than experiment were the quantum oscillations were to arise from a surface atomic layer.

We look for experimental evidence for low energy excitations within the charge gap of $SmB_6$, which we compare with the Fermi surface associated with the measured quantum oscillations. The inset of Fig. 2a shows specific heat capacity measured in multiple single crystals of $SmB_6$ grown by the floating zone technique and by the flux growth technique, which are either the same samples on which quantum oscillations are observed, or from the same crystal growth. At low temperatures, a finite density of states at the Fermi energy within the charge gap is revealed by a finite value of the linear specific heat coefficient $\gamma \approx 4(2)$ mJ·mol$^{-1}$·K$^{-2}$ (see Methods), which rapidly increases with decreasing temperature (Fig. 2a). We compare the size of the density of states measured from the linear specific heat coefficient, with the expectation from the three-dimensional ellipsoidal Fermi surface geometry we fit to the measured quantum oscillations (shown in Fig.1b), and effective masses measured from quantum oscillations (see figure in ref. [17] and Fig. 2c). A common origin of in-gap low energy excitations is indicated from the good agreement we find between the value of density of states calculated from the quantum oscillation-extracted Fermi surface $\gamma \approx 4(1)$ mJ·mol$^{-1}$·K$^{-2}$ on assuming a contribution from the entire sample volume, and the value measured from the linear specific heat capacity (Fig. 2c, Methods and ref. [18]). Further, the steep increase at low temperatures of the value of $\gamma$ closely resembles the steep increase of quantum oscillation amplitude at low temperatures observed for the majority of quantum oscillation frequencies in the case of floating zone-grown SmB$_6$ (Figs. 2a-b, see figure in ref. [17]) [13]. The nuclear contribution to the specific heat capacity is expected to be negligibly small in the experimental temperature range below 1 K at zero magnetic field [19, 20] (see Methods), although an increase in nuclear contribution in the presence of a magnetic field makes the accurate determination of low temperature linear specific heat capacity in finite applied magnetic fields challenging (see Methods) [21].

Another probe of itinerant in-gap low energy excitations is provided by an experimental estimate of the low temperature quantum oscillatory component of the entropy. We estimate this low temperature entropy by measuring $dM_{osc.}/dT$ (where $M_{osc.}$ is the quantum oscillatory magnetisation), which is related to the entropy by the Maxwell relation $V\left(\frac{\partial M}{\partial T}\right)_B = \left(\frac{\partial S}{\partial B}\right)_T$, where $V$ is the volume of the crystal, $M$ is the magnetisation, and $S$ is the entropy (see Methods). The value of $dM_{osc.}/dT$ (shown in Fig. 2d) remains finite in amplitude down to temperatures $< 1$ K, an order of magnitude below the charge gap 2-5 meV [8, 13, 21], providing evidence for a finite density of states within the charge gap. Importantly, we are able to demonstrate the itinerant character of the in-gap density of states since the accessed entropy is oscillatory, derived from the measured oscillatory magnetisation.

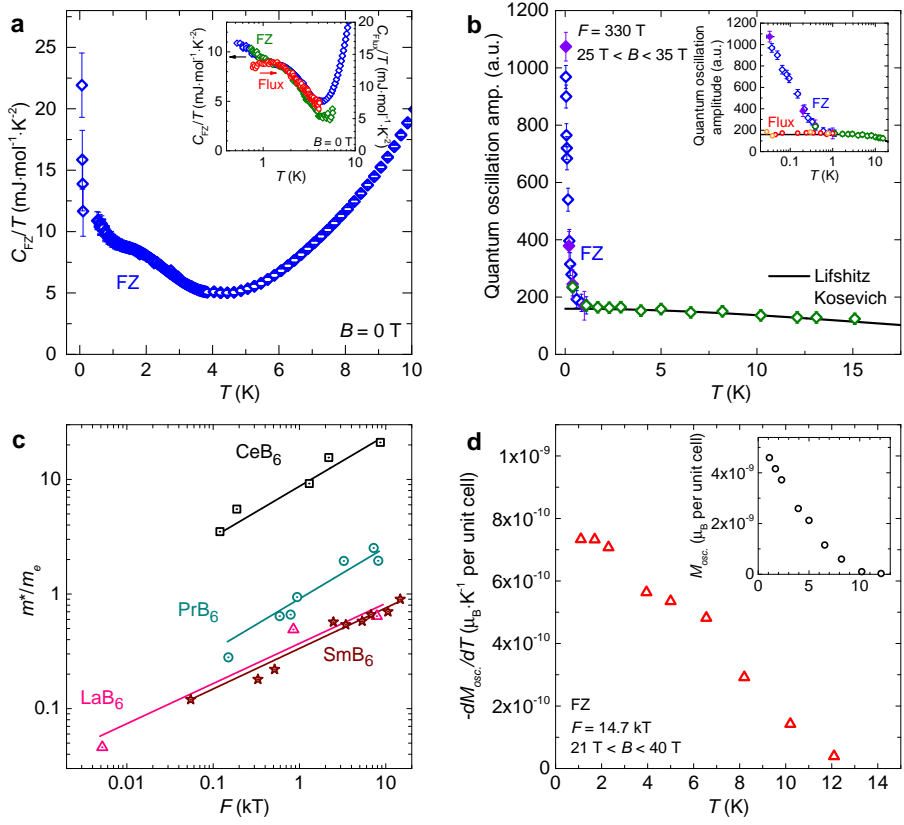

Figure 2: **Finite linear specific heat coefficient and quantum oscillatory entropy of SmB$_6$.**
**a**, Measured specific heat capacity of SmB$_6$ for a floating zone-grown single crystal (FZ) down to 63 mK, revealing a finite heat capacity divided by temperature at low temperatures, unexpected for an insulator, with a surprising steep increase below $\approx 1$ K (similar to ref. [18]). The inset shows the measured specific heat capacity for two floating zone-grown crystals (blue and green diamonds) and a flux-grown crystal (circles), demonstrating a finite heat capacity divided by temperature at low temperatures across all samples, with a more prominent increase below $\approx 1$ K exhibited by floating zone-grown crystals. **b**, Steep non Lifshitz-Kosevich low temperature upturn below $\approx 1$ K in quantum oscillation amplitude for the case of floating zone-grown SmB$_6$ (purple, blue, green diamonds, with the error corresponding to the noise floor of the Fourier transform), with similarities to the low temperature upturn in the heat capacity divided by temperature which is most prominent for this type of sample, shown in **a**. The inset shows the prominent increase in quantum oscillation amplitude below approximately 1 K that deviates from Lifshitz-Kosevich temperature dependence is only observed in the case of floating zone-grown crystals (three samples shown by purple, blue, and green diamonds), and not flux-grown crystals (two samples shown by orange and red diamonds). **c**, Measured effective mass of the various frequency branches of SmB$_6$ from a Lifshitz-Kosevich fit down to 1 K (star symbols, see figure in ref. [17]), seen to be very similar to the metallic rare-earth hexaborides [14, 15, 16, 54], especially nonmagnetic LaB$_6$. (Next page.)

Figure 2: (Previous page.) **d**, Derivative with respect to temperature of the highest frequency ($\alpha$) magnetic quantum oscillation amplitude remains finite to low temperatures, reflecting a finite quantum oscillatory entropy at temperatures well below the transport gap scale (see Methods). The inset shows the magnetic quantum oscillation amplitude of the $\alpha$ frequency as a function of temperature down to $\approx$ 1 K in a floating zone-grown sample.

A further test of the itinerant nature of measured bulk in-gap low energy excitations is provided by a measurement of the thermal conductivity at temperatures $\ll$ 1 K, where the phonon contribution is strongly suppressed. Fig. 3a shows the measured low temperature thermal conductivity of a single crystal of $SmB_6$ grown using the floating zone technique. The phonon contribution up to high temperatures can be modelled well by boundary limited phonons, shown by the red line denoted by $\kappa_{ph.}/T$, accounting for the zero field thermal conductivity, and is characteristic of high sample quality (see Methods). On subtracting the phonon contribution from the measured thermal conductivity (inset to Fig. 3a), the remainder is seen to be very small in zero field, but becomes increasingly large in an applied magnetic field, far exceeding the Wiedemann-Franz expectation from the surface conducting layer by orders of magnitude (see Methods). An origin of this additional contribution from phonons is unlikely, since the phonon thermal conductivity is already at a maximum in the boundary scattering limit. The possibility of a conventional magnon contribution is also not supported due to the absence of static magnetic moments as inferred from muon spin resonance measurements [22], neutron scattering measurements [23], and magnetisation measurements (see figure in ref. [17]).

Intriguingly, a similar observation of a substantial enhancement in low temperature thermal conductivity with applied magnetic field has been observed in the Mott insulating organic systems EtMe$_3$Sb[Pd(dmit)$_2$]$_2$ and $\kappa$-(BEDT-TTF)$_2$Cu$_2$(CN)$_3$ [24, 25, 26, 27] (shown in Figs. 3c-d), which have been associated with a theoretical model of novel spinon low energy excitations that transport heat but not charge [28, 29, 30, 31]. Both systems display a finite linear specific heat capacity coefficient, while in EtMe$_3$Sb[Pd(dmit)$_2$]$_2$ the thermal conductivity displays a finite linear temperature dependence at low temperatures, in $\kappa$-(BEDT-TTF)$_2$Cu$_2$(CN)$_3$ the thermal conductivity displays a downturn as a function of temperature at millikelvin temperatures. These experimental observations were collectively interpreted in terms of a neutral Fermi surface in the organic spin liquid materials, potentially evincing a low temperature instability in $\kappa$-(BEDT-TTF)$_2$Cu$_2$(CN)$_3$. The intriguing similarity of our observations in SmB$_6$ points to a neutral Fermi surface comprising itinerant low energy excitations that transport heat, but not charge in SmB$_6$. Informing the search for more examples of similar material systems, we note that such experimental signatures of neutral low energy excitations are likely to be more prominent in materials positioned closer to gaplessness of neutral low energy excitations, potentially tuned by factors such as applied magnetic field and materials parameters (Fig. 4).

A sufficiently large effective mean free path of itinerant low energy excitations is important for the observation of magnetic quantum oscillations, thermal conductivity, and quantum oscillatory entropy, in contrast to the measured specific heat capacity. A comparison between measured quantities into which the effective mean free path enters is most meaningful at high magnetic fields, where high frequency quantum oscillations corresponding to the largest ellipsoidal ($\alpha$) Fermi surface that dominates the density-of-states at the Fermi energy are observable. Using assumptions relevant to a conventional metal with electronic excitations, the value of excess thermal conductivity we measure in floating zone-grown samples of SmB$_6$ in an applied magnetic field of 12 T and at temperatures of $\approx 200$ mK corresponds to a mean free path estimate of the dominant large Fermi surface of $\approx 10^{-8}$ m. This estimate is similar to the estimated mean free path of a few times $10^{-8}$ m obtained from the measured cyclotron radius and exponential damping (Dingle) term from quantum oscillations in magnetic fields of $35-45$ T in floating zone-grown samples (see Methods). The significantly larger exponential damping term that renders the high frequency oscillations considerably smaller in size for the flux-grown samples compared to the floating zone-grown samples (see Methods), is consistent with the lower magnetic field enhancement of the thermal conductivity seen for these samples (Fig. 3b inset). A three-dimensional Fermi surface associated with bulk in-gap itinerant low energy excitations in $SmB_6$ is thus supported by our collective measurements down to low temperatures of specific heat, magnetic quantum oscillations, thermal conductivity, and quantum oscillatory entropy. Recent nuclear magnetic resonance (NMR) measurements also reveal consistent signatures of an NMR relaxation rate divided by temperature which is constant as a function of temperature at low temperature, instead of exponentially vanishing, as would be expected for a gapped density of states ([19] and unpublished).

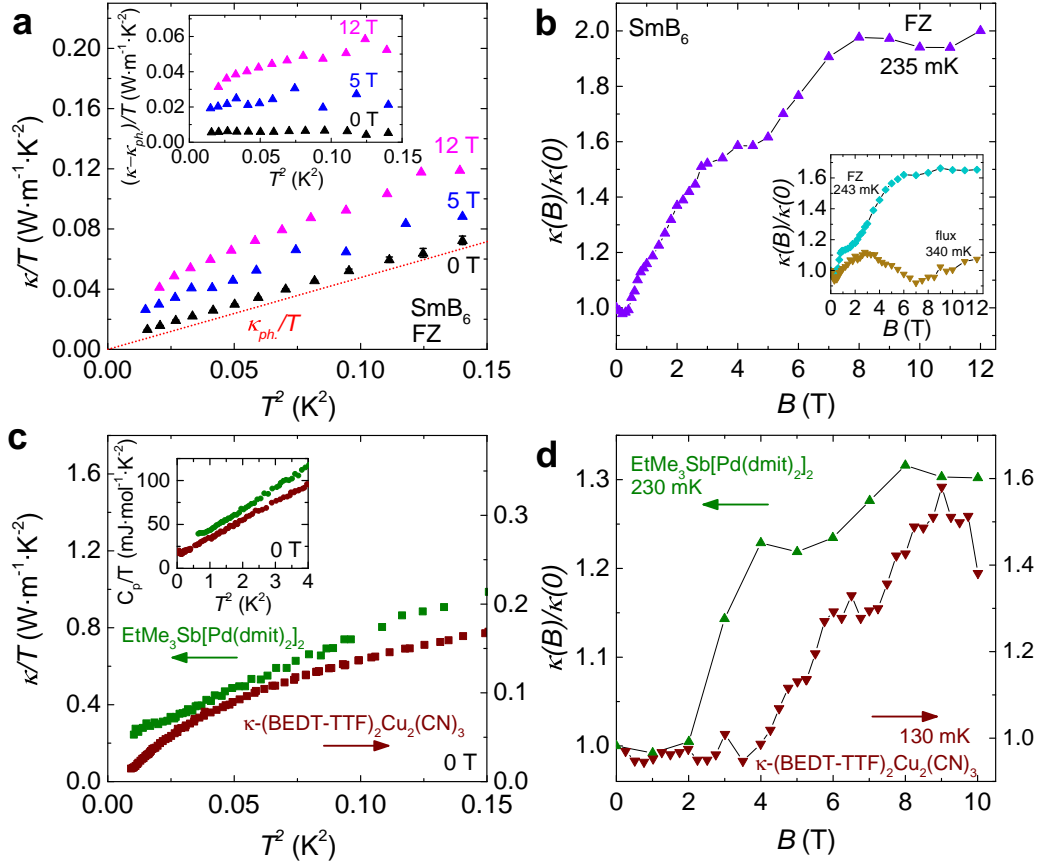

Figure 3: **Low temperature thermal conductivity of SmB$_6$.** **a**, Thermal conductivity ($\kappa$) of a floating zone-grown single crystal of SmB$_6$ plotted as $\kappa/T$ as a function of $T^2$. The value of thermal conductivity at the lowest measured temperature shows a nearly fourfold increase in increasing applied magnetic fields up to 12 T. The zero field data is largely accounted for by the phonon contribution (see Methods) calculated for a Debye temperature of $\Theta_D = 373$ K (red line denoted by $\kappa_{ph.}/T$), obtained from elastic constants [55]. The enhancement in a magnetic field is clearly seen in the inset upon subtracting the phonon contribution. The thermal gradient is applied along the [100] direction, with perpendicular magnetic field applied along [001]. **b**, Thermal conductivity as a function of magnetic field shows a significant increase with magnetic field for a floating zone-grown single crystal. The inset shows a similarly large increase with magnetic field for a second floating zone-grown single crystal, while the enhancement for a flux-grown crystal is subtle [37]. **c**, Low temperature thermal conductivity measured on two different organic insulating spin liquids, taken from ref. [24], both of them associated with a finite linear specific heat coefficient (inset [26, 27]), resembling our findings in SmB$_6$. **d**, Large magnetic field dependence of the low temperature thermal conductivity measured in both organic spin liquids (from refs. [24, 25]), is seen to be remarkably similar to our measurements in floating zone-grown SmB$_6$.

Our experimental results appear inconsistent with theoretical models that do not involve a bulk in-gap density of states, such as those that invoke for instance surface states, quenched disorder or interband tunneling phenomena [12, 32, 33, 34, 35, 36, 37]. We consider various proposed alternative theoretical models that invoke novel itinerant low energy excitations within the charge gap in $SmB_6$ [28, 29, 30, 31, 38, 39, 40, 41, 42, 43, 44], including magnetic excitons [38], neutral quasiparticles such as spinons [28, 29, 30, 31], composite excitons [41] and Majorana fermions [42, 43, 44], and compare them with our key experimental observations. A more extensive compilation of theoretical models proposed to explain quantum oscillations in $SmB_6$ is provided in the methods section.

A spinon model [28, 29, 30, 31] was earlier proposed for a single band Mott insulating organic spin liquid, in which case a spinon Fermi surface arises from these neutral fermionic particles. In this model, diamagnetism arises from the effects of non bilinear terms in the spin Hamiltonian that depend on the applied magnetic field. The meaning of such field dependent terms can be understood in a higher energy description that includes virtual charge fluctua-tions over an extended range of sites [31, 45], the amplitude of which is enhanced close to the insulator-metal phase boundary. Coupling to the electric field vanishes in the DC limit, but is predicted to be finite in the finite frequency limit. This prediction is consistent with the observation of substantial bulk conductivity in $SmB_6$ at a frequency of a few hundred GHz evidenced by time domain terahertz spectroscopic experiments [46]. Caveats to this model include the suggestion that quantum oscillations might not be observed in practice in the case of a single-band Mott insulator due to the formation of Condon domains [30]. It is also unclear as to the quantum oscillation frequencies that would be observed, given the potential difference between the effective and applied magnetic field in this model [30]. In order to further probe such a scenario, experiments to search for low energy spin excitations are indicated to complement the high energy collective mode at 14 meV seen through inelastic neutron scattering [23], which is at too high an energy scale to be directly related to the phenomena we observe.

More recently a magnetic exciton model [38] has been proposed, within which the low energy excitations are bosonic in character. Instead, fermionic excitations are associated with a composite exciton model [41], which has recently been proposed for a strongly correlated three-dimensional mixed valence insulator in the limit of strong Coulomb interaction. Under suitable conditions a collective state of neutral fermionic composite excitons is predicted, which would yield a Fermi surface of the same volume as the original conduction $d$-electron Fermi surface, similar to our observations. A finite linear specific heat coefficient, a finite thermal conductivity divided by temperature, a constant NMR relaxation rate divided by temperature at low temperatures, and appreciable frequency-dependent optical conductivity are predicted, in agreement with our findings and other experiments [19, 46]. Quantum oscillations of the free energy periodic in the inverse internal magnetic field are also predicted [47], although it is not clear as to the size of the effective magnetic field that would be felt by the composite excitons compared to the size of the physical applied magnetic field. In addition to quantum oscillations in the magnetisation, quantum oscillations in the electrical resistivity are also predicted to appear for materials positioned closer to the insulator metal transition [47]. The observation of a finite bulk thermal Hall effect would further establish a strong correspondence between the effective magnetic field felt by the composite exciton, and the physical applied magnetic field within this model.

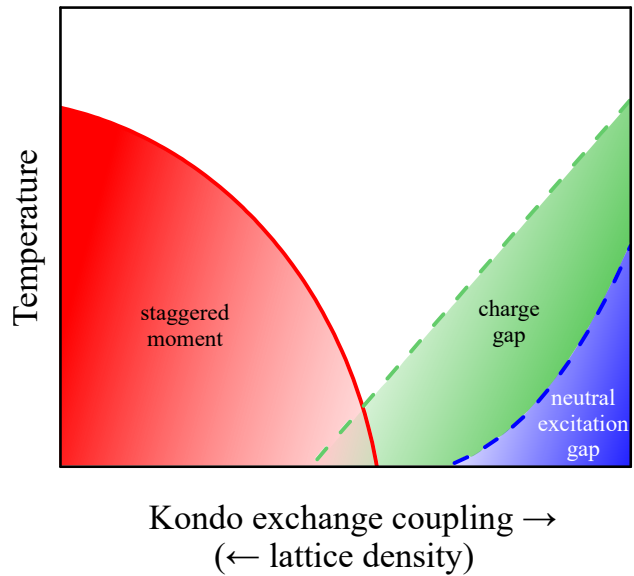

Figure 4: **Schematic phase diagram adapted from numerical simulations**. Phase diagram adapted from Monte Carlo simulations of a magnetic Kondo lattice model [56], which indicate a collapse of the neutral low energy gap in the region where the charge gap is still finite. Our measurements suggest the location of $SmB_6$ in the region of a small finite charge gap, but on the brink of gapless neutral low energy excitations. More prominent experimental signatures of neutral low energy excitations are likely to be observed in materials positioned even closer to gaplessness, potentially tuned by external variables such as applied magnetic field, or for $SmB_6$ - increasing lattice density, as well as other materials parameters.

We next consider the Majorana fermion model [42, 43, 44], where in contrast to the better known slave-boson mean field model, the coupling of doubly degenerate conduction and $f$-electron bands leads to four Majorana bands, one of which coincides in energy with the starting conduction band but represents the spectrum of neutral rather than charged excitations. Within this model, a Fermi surface of Majorana fermions therefore corresponds to the conduction electron Fermi surface (i.e. the same as the Fermi surface of $R\text{B}_6$), in agreement with experiment. While the electric current vanishes to lowest order in applied electric field in this model, it is expected to be finite to second order, yielding a diamagnetic response. A frequency-dependent optical conductivity response is further predicted, in agreement with time domain terahertz spectroscopic experiments [46]. The ground state of this model is predicted to be a triplet superconductor in which long range order is destroyed by fluctuations [44], the amplitude of which is predicted to be magnetic field dependent, yielding a linear increase in low temperature thermal conductivity in qualitative agreement with our experimental observation (Fig. 3b). Further predictions of this model, such as the appearance of a superconducting Meissner effect at low temperatures and low magnetic fields, remain to be further experimentally investigated [44].

The salient findings that identify a Fermi surface of neutral low energy excitations within the charge gap are common to classes of samples grown by different techniques, as these exhibit essentially the same specific heat capacities and bulk quantum oscillations in the magnetization above 1 K (Fig. 1, insets to Fig. 2a,b). The Fermi surface and quasiparticle effective masses inferred from these oscillations are consistent with the measured coefficient of the linear heat capacity (Figs. 1, 2a-c). Moreover, the oscillatory entropy inferred from the temperature derivative of the oscillatory magnetisation (Fig. 2d) confirms the itinerant nature of the excitations within the charge gap. Differences below 1 K in observed quantities (seen in insets to Figs. 2a,b and to 3b) do not affect our above key conclusions, and are likely due to subtle materials property differences due to different growth conditions [48]. Similar sensitivity to preparation technique has been reported, for example, in the classic heavy fermion superconductor $CeCu_2Si_2$, in which case the sensitivity has been interpreted in terms of effects such as differing lattice density in samples prepared by different techniques [49].

Theoretical models of a Fermi surface from neutral quasiparticles are suggested as an explanation for the breadth of surprising experimental observations in Kondo insulating $SmB_6$, although quantitative comparisons especially with the size of measured quantum oscillations, remain outstanding. The physics captured by these mean field models may be similar to a dynamic model invoking slow fluctuations between a collectively hybridised insulating state and an unhybridised dynamic state with a Fermi surface of conduction electrons [13]. We note that our analysis of the experimental data and theoretical models proposed thus far assume a description in terms of low energy excitations. An outstanding possibility is the need for a description that transcends quasiparticles, such as new classes of topological models [50] and holographic models [51]. Our work has identified a new route for the realisation of the landmark paradigm of a Fermi surface in the absence of a Fermi liquid in the class of Kondo insulators positioned at the brink of a Kondo insulator to metal transition. Similar experiments are indicated to search for clues in other families of Kondo insulators, including $YbB_{12}$ [52], the system most similar to $SmB_6$ with a comparable size of charge gap and a finite measured linear specific heat capacity, and SmS [53], which provides tuning possibilities to approach the insulator metal transition via applied pressure.

## Acknowledgements

M.H., Y.-T.H., G.R.-G., J.B., M.K.K., G.H.Z., and S.E.S. acknowledge support from the Royal Society, the Winton Programme for the Physics of Sustainability, EPSRC (studentship and grant number EP/P024947/1) and the European Research Council under the European Unions Seventh Framework Programme (grant number FP/2007-2013)/ERC Grant Agreement number 337425. S.E.S. acknowledges support from the Leverhulme Trust by way of the award of a Philip Leverhulme Prize. Work done by W.H.T. and R.W.H. was funded by NSERC of Canada. Q.R.Z., B.Z. and L.B. acknowledge support from DOE-BES through award DE-SC0002613. X.C. and M.S. acknowledge support from Corpus Christi College, Cambridge and EPSRC. M.C.H. and G.B. would like to acknowledge financial support from the EPSRC, UK through Grants EP/M028771/1 and EP/L014963/1. Work done by S.N. and T.S. was supported by a Grant-in-Aid for Scientific Research on Innovative Areas "J-Physics" (15H05883) and KAKENHI (15H03682) from MEXT. M.K.C. and N.H. acknowledge support from the US Department of Energy, Office of Science, BESMSE Science of 100 Tesla program. S.Y. acknowledges support from Grant-in-Aid for Scientific Research JP16K05447. G.G.L. acknowledges support from EPSRC grant EP/K012894/1. A portion of this work was performed at the National High Magnetic Field Laboratory, which is supported by National Science Foundation Cooperative Agreement No. DMR-1157490, the State of Florida and the DOE. We acknowledge discussions with many colleagues, including M. Aronson, G. Baskaran, P.-Y. Chang, D. Chowdhury, P. Coleman, N. R. Cooper, M. P. M. Dean, O. Erten, J. Flouquet, J. Knolle, N. J. Laurita, P. A. Lee, P. B. Littlewood, V. F. Mitrović, J. E. Moore, T. P. Murphy, M. Norman, C. Pépin, S. Sachdev, T. Senthil, Q. Si, A. M. Tsvelik, C. Varma. We are grateful for the experimental support provided by the NHMFL, Tallahassee, including J. Billings, R. Carrier, E. S. Choi, B. L. Dalton, D. Freeman, L. J. Gordon, M. Hicks, S. A. Maier, J. N. Piotrowski, J. A. Powell, E. Stiers.

## Methods

**Conversion of measured quantum oscillations into bulk magnetic moment per unit cell.**
Magnetic torque was measured via the capacitive torque technique, with a typical oscillation size of $\approx 4 \cdot 10^{-4}$ pF in the measured capacitance at a magnetic field of 15 T (see figure in ref. [17]). Using the dimensions and Young's modulus of our cantilever, we obtained a spring constant $k = 28(8)$ N·m$^{-1}$. Similar values were found by estimation from displacement under gravity, and displacement under a magnetic field gradient (Faraday balance). The torque $\tau$ on the cantilever is proportional to its deflection, given by $\tau = Lk\delta$, where $L$ is the length of the cantilever, and $\delta$ is the deflection, which is in turn proportional to the change in capacitance by $\delta = d_0 \cdot \Delta C / C$, with $d_0$ being the distance between the opposing faces of the cantilever and the bottom plate. The torque is related to the total magnetic moment $\mu$ via $\tau = \mu B \sin \theta_M$, where $\theta_M$ is the angle between the magnetic field $B$ and the total magnetic moment $\mu$. We express the magnetic moment $p_s$ in units of Bohr magnetons per unit cell, by writing $\mu = (s/a_{\mathrm{u.c.}})^3 p_s \mu_B$, where $s^3$ is the volume of the crystal, and $a_{\mathrm{u.c.}}$ is the lattice constant. Our final expression is therefore

$$\Delta p_s = \frac{d_0 L k a_{\mathrm{u.c.}}^3}{s^3 \mu_B B C \sin \theta_M} \cdot \Delta C \qquad (1)$$

Using $d_0 = 0.1$ mm, $L = 3.8$ mm, $k = 28$ N·m$^{-1}$, $a_{\mathrm{u.c.}} = 0.413$ nm, $s^3 = 0.5 \cdot 0.8 \cdot 0.4$ mm$^3$, this becomes

$$\Delta p_s = \frac{0.51}{B \sin \theta_M} \cdot \Delta C \ \mathrm{T \cdot pF^{-1}} \ \mu_\mathrm{B} \ \text{per unit cell} \qquad (2)$$

for the SmB$_6$ measurements. From Fig. 1e we estimate the amplitude (zero to peak) of the oscillations to be $\approx \frac{1.1 \cdot 10^{-5}}{\sin \theta_M} \mu_B$ per unit cell at $B = 18$ T. Here, $0.1 \lesssim \sin \theta_M \lesssim 1$ depending on the orientation of the magnetic moment.

For LaB$_6$, using a cantilever with slightly different dimensions, we have $d_0 = 0.1$ mm,

$L = 3.8$ mm $k = 17(5)$ N·m$^{-1}$, $a_{\text{u.c.}} = 0.416$ nm, $s^3 = 1.0 \cdot 1.0 \cdot 0.25$ mm$^3$, and therefore

$$\Delta p_s = \frac{0.20}{B \sin \theta_M} \cdot \Delta C \text{ T} \cdot \text{pF}^{-1} \; \mu_\text{B} \text{ per unit cell} \tag{3}$$

From Fig. 1d we find the amplitude of the oscillations to be $\approx \frac{1.3 \cdot 10^{-5}}{\sin \theta_M}$ $\mu_\text{B}$ per unit cell at $B = 9$ T. Here, $0.1 \lesssim \sin \theta_M \lesssim 1$ depending on the orientation of the magnetic moment (the angle is taken to be positive throughout).

**Calculation of the theoretical amplitude of bulk de Haas-van Alphen oscillations.** The fundamental oscillatory magnetisation $M$ in the Lifshitz-Kosevich theory is given by

$$M = D \cdot R_T R_D R_S \cdot \sin(2\pi F/B + \phi) \tag{4}$$

where $R_T$, $R_D$, and $R_S$ are the usual damping terms due to finite temperature, scattering, and spin-splitting (see, e.g., ref. [57] and [58]). The exponential damping term $R_D$ is expressed as $R_D = \exp(-B_0/B)$, where $B_0$ reflects the strength of damping of the quantum oscillation amplitude for each sample and frequency. $D$ is the infinite field, zero spin-splitting amplitude given by

$$D = -\frac{\mu_B A_F^{3/2} m_e}{2\pi^4 m^*} \sqrt{\frac{B}{F|A''|}} \tag{5}$$

where $\mu_B$ is the Bohr magneton, $A_F$ is the Fermi surface area normal to the magnetic field $B$, $m^*$ is the effective mass in absolute units, $F$ is the oscillation frequency, and $|A''|$ is the second derivative of the Fermi surface area with respect to the effective wave vector along $B$. We can define the moment per unit cell in units of Bohr magnetons as $Dv/\mu_B$, where $v = a_{\text{u.c.}}^3$ is the volume of the unit cell, so that the peak amplitude in the infinite field and zero spin-splitting limit is

$$p_s = \frac{|D|v}{\mu_B} = \sqrt{\frac{2\pi}{|A''|}} \frac{m_e}{m^*} \left(\frac{a_{\text{u.c.}} k_F}{\pi}\right)^3 \sqrt{\frac{B}{8F}} \tag{6}$$

where we define $k_F$, the effective Fermi wave vector, via $A_F = \pi k_F^2$. The anisotropy term, $\sqrt{2\pi/|A''|}$, is dependent on the eccentricity $r$ of the ellipsoidal Fermi surface, and hereafter will be written as $f(r)$.

**Comparison of quantum oscillation amplitude in SmB$_6$ and LaB$_6$ with theoretical amplitude.**

The comparable size of quantum oscillations in the infinite field quantum limit measured in SmB$_6$ and LaB$_6$ is shown in Fig. 1 as a function of the phase $F/B$. For the lowest frequency $\rho'$ branch in SmB$_6$, the experimentally measured values correspond to $F = 31$ T, $m^*/m_e = 0.12$, $a_{\text{u.c.}} = 0.413$ nm, and $R_D = \exp(-30\text{ T}/B)$ as inferred from Fig. 1e. Estimating $f(r) \approx 1$-2, $R_S = 0.5$-1, and taking into account a degeneracy factor of 2-8, the expectation for the theoretical amplitude of the magnetic moment for the $\rho'$ frequency branch of SmB$_6$ is of the order $\approx 10^{-5}$-$10^{-4}$ $\mu_B$ per unit cell at $B = 16.7$ T, including the angular anisotropy term $f(r)$, Dingle $R_D$ and spin-splitting $R_S$ damping factors. This is consistent in order of magnitude with the experimentally measured amplitude of quantum oscillations shown below Eq. 2. Similarly for LaB$_6$, the low frequency oscillations correspond to experimentally measured values $F = 5.9$ T, $m^*/m_e = 0.05$, $a_{\text{u.c.}} = 0.416$ nm, $R_D = \exp(-1\text{ T}/B)$ as inferred from Fig. 1d. Estimating $f(r) \approx 1$-2, $R_S = 0.5$-1, and taking into account a degeneracy factor of 2 (from ref. [15]), we find the theoretical amplitude to be of order $\approx 10^{-4}$ $\mu_B$ per unit cell at $B = 6.2$ T, including the angular anisotropy term $f(r)$, Dingle $R_D$ and spin-splitting $R_S$ factors, again consistent with the measured value shown below Eq. 3.

In Fig. 1g, the theoretically predicted amplitude of quantum oscillations in magnetic torque (M × B where $\theta_M$ is the angle between M and B) rather than magnetisation (M) is plotted, where M is in units of $\mu_B$ per unit cell, and $B$ is in units of tesla. Given the range of $0.1 \lesssim \sin\theta_M \lesssim 1$, we use an intermediate value of $\sin\theta_M \approx 0.5$ for the simulation in Fig. 1g. Intermediate values are also used of $R_D \approx 0.166$, $R_S \approx 0.75$ for both the three-dimensional and two-dimensional simulation, as well as $f(r) \approx 4$, and a degeneracy factor of 4 for the three-dimensional simulation in Fig. 1g.

The exponential damping term in the case of SmB$_6$ is considerably higher than in LaB$_6$, as indicated from the magnetic field dependence shown in the figure provided in ref. [17], which reveals a higher onset in magnetic field of observable quantum oscillations in SmB$_6$ compared to metallic LaB$_6$. Both high magnetic fields and extremely high experimental sensitivity are thus required to access especially high frequencies in SmB$_6$. We note that while samples of SmB$_6$ prepared by different techniques yield the same quantum oscillation frequencies, sizeable variations can occur in the measured quantum oscillation amplitude; samples with the largest quantum oscillation amplitude are selected for study on account of their high inverse residual resistivity ratio and low finite specific heat coefficient, and by extensive screening in high magnetic fields.

**Comparison with surface quantum oscillation model for SmB$_6$.** The theoretical quantum oscillation size is obtained from the carrier density corresponding to a two dimensional cylindrical Fermi surface. In the two-dimensional limit, the carrier density is directly related to the Fermi surface area. Hence for the small ellipsoidal pockets that occupy a tiny fraction of the Brillouin zone (the volume of the $\rho'$ pockets constitute 0.1% of the Brillouin zone), the theoretical amplitude of quantum oscillations is expected to be very small. The carrier density per unit surface area is given by

$$n = \frac{2}{(2\pi)^2} \pi k_F^2 \tag{7}$$

including a factor of 2 for spin degeneracy. For the lowest observed quantum oscillation frequency of $F = 31$ T, we find $n = 1.5 \cdot 10^{16}$ m$^{-2}$. Defining the moment per unit cell in units of Bohr magneton, the peak amplitude in the infinite field and zero spin-splitting limit is

$$p_s = n a_{u.c.}^2 \frac{2m_e}{\pi m^*} = \frac{4m_e}{m^*} \left(\frac{k_F}{k_{BZ}}\right)^2 \tag{8}$$

where $k_{\rm BZ} = 2\pi/a_{u.c.}$, and $m^*$ is the effective mass in absolute units. The peak amplitude of the quantum oscillations is found to have a theoretical maximum value of $\approx 10^{-2}$ $\mu_B$ per surface unit cell in the infinite field limit prior to including Dingle and spin-splitting damping terms, which would reduce the theoretically predicted value to $\approx 10^{-3}$ $\mu_B$ per surface unit cell at 18 T. In contrast, the measured quantum oscillations would correspond to an extremely large magnetic moment per surface unit cell were they to arise from the surface, given that the surface unit cells constitute only a tiny fraction $\sim 10^{-6}$ of the total number of unit cells. The measured peak amplitude of the quantum oscillations on considering a surface origin would correspond to a magnetic moment per surface unit cell of at least $\approx 10$ $\mu_B$ per surface unit cell at 18 T, a value which would be even larger on accounting for the orientation of the magnetic moment $\sin\theta_M$ (Eq. 2). Such a large value is several orders of magnitude larger than the theoretical maximum quantum oscillation size predicted for a surface atomic layer origin, ruling out such an origin as an explanation for the quantum oscillations reported here. The high values reported for the low-frequency quantum oscillations in ref. [12] are also at least an order of magnitude larger than the theoretical maximum.

**Quantitative comparison of the density of states at the Fermi energy from measured linear specific heat coefficient and from measured quantum oscillations.** Within the traditional Fermi liquid theory, the quasiparticle density of states at the Fermi energy is directly related to the linear specific heat coefficient $\gamma$ by

$$N(E_F) = \frac{3\gamma}{\pi^2 k_B^2} \qquad (9)$$

We compare the quasiparticle density of states corresponding to the measured linear specific heat capacity coefficient with that corresponding to the Fermi surface measured from quantum oscillations. For a known Fermi surface geometry and quasiparticle velocity, the quasiparticle density of states at the Fermi energy is given by

$$N(E_F) = \frac{1}{4\pi^3 \hbar} \int_S \frac{dS}{|v^*|} \qquad (10)$$

After Fig. 1, the main Fermi surface features in $SmB_6$ can be described by ellipsoidal electron sheets, similar to other rare earth hexaborides. Ellipsoids with semi-principal axes $ak_0$, $bk_0$ and $ck_0$ can be parametrised by

$$E_F = \frac{\hbar^2 k_x^2}{2a^2 m^*} + \frac{\hbar^2 k_y^2}{2b^2 m^*} + \frac{\hbar^2 k_z^2}{2c^2 m^*} \qquad (11)$$

with $k_x = ak_0 \cos\phi \sin\theta$, $k_y = bk_0 \cos\phi \cos\theta$ and $k_z = ck_0 \sin\phi$, $k_0$ is a constant, and $a$, $b$ and $c$ represent the relative ratios of the semi-principal axes. The area element in the integral becomes that of an ellipsoid:

$$dS = k_0 \cdot \cos\phi \sqrt{a^2 b^2 \sin^2\phi + c^2 \cos^2\phi (a^2 \sin^2\theta + b^2 \cos^2\theta)}\, d\phi d\theta \qquad (12)$$

A full description of the quasiparticle velocity $v^*$ can be obtained for the Fermi surface described by Eq. 11, via

$$|v^*| = |(1/\hbar)\nabla_k E_F| = \frac{\hbar k_0}{abcm^*}\sqrt{a^2 b^2 \sin^2\phi + c^2 \cos^2\phi (b^2 \sin^2\theta + a^2 \cos^2\theta)} \qquad (13)$$

These allow the integral in Eq. 10 to be carried out over $\phi$ from $-\pi/2$ to $\pi/2$, and $\theta$ from 0 to $2\pi$, to obtain the density of states, which can be computed for known semi-principal axes and effective mass. In the special case of prolate ellipsoids this would lead to the result obtained in ref. [59]. Here we assume contribution from both spin up and spin down Fermi surfaces.

The table provided in ref. [17] shows the effective masses and semi-principal axes obtained for each Fermi surface sheet. Their contribution to the linear specific heat coefficient $\gamma$ adds up to $\gamma = 4(1)$ mJ·mol$^{-1}$·K$^{-2}$, with the large $\alpha$ sheet contributing 3 mJ·mol$^{-1}$·K$^{-2}$. LaB$_6$ has a comparable $\alpha$ sheet [14], giving $\gamma = 2.6$ mJ·mol$^{-1}$·K$^{-2}$ following this calculation, the same value as found by ref. [59]. This is smaller than the contribution from the $\alpha$ sheet found for SmB$_6$ due to the smaller eccentricity and effective mass.

**Specific heat capacity measured for multiple samples and in a magnetic field.** The measured linear specific heat coefficient is found to be similar for all samples studied in this work. The range of values of the linear specific heat coefficient presented in the main text ($\gamma \approx 4(2)$ mJ·mol$^{-1}$·K$^{-2}$) reflects three different samples (two floating zone- and one flux-grown) after phonon subtraction. The larger linear specific heat capacity reported in refs. [6, 60] is most likely due to an additional contribution from a small fraction of Sm vacancies, as inferred from Raman spectroscopy measurements [61].

The figure provided in ref. [17] shows specific heat measured in magnetic fields up to 14 T and down to 700 mK for floating zone-grown SmB$_6$. We find that the specific heat in an applied magnetic field retains the features seen at zero magnetic field. We note, however, as also observed in ref. [21], that the increase in nuclear contribution with magnetic field at low temperatures can make the extraction of linear specific heat in a magnetic field challenging, as the nuclear contribution to the specific heat is proportional to the square of the magnetic field and inversely proportional to the square of the temperature [62].

**Negligible nuclear contribution to the heat capacity in zero magnetic field.** In SmB$_6$, only boron contributes to the nuclear quadrupole specific heat $C_Q$, because the samarium site in the crystal has a cubic symmetry and hence the electric field gradient is zero. Boron has two isotopes, $^{10}$B (natural abundance $x_{10} = 19.9\%$, nuclear spin $I_{10} = 3$, nuclear electric quadrupole moment $Q_{10} = 84.6$ millibarn) and $^{11}$B ($x_{11} = 80.1\%$, $I_{11} = 3/2$, $Q_{11} = 40.6$ millibarn). In zero magnetic field, the $^{10}$B spin has four energy levels owing to the electric field gradient [63], resulting in a four-level scheme that gives the expression $C_{10} = 12R(h\nu_{10}/k_BT)^2$ per mole of $^{10}$B for the specific heat for $k_BT \gg h\nu_{10}$. The $^{11}$B spin has two energy levels in zero magnetic field, with a specific heat of $C_{11} = R/4(h\nu_{11}/k_BT)^2$ per mole at $k_BT \gg h\nu_{11}$. Here, $\nu_{10}$ and $\nu_{11}$ represent the nuclear quadrupole resonance frequencies of $^{10}$B and $^{11}$B. The nuclear quadrupole resonance frequency of $^{11}$B has been measured in SmB$_6$ by several groups [19, 63], giving $\nu_{11} = 0.570$ MHz. $\nu_{10}$ can be estimated from $\nu_{11}$, as the nuclear quadrupole resonance frequency is given by $\nu_Q = 3eQV_{zz}/[2hI(2I-1)]$, where $V_{zz}$ is the largest principal axis component of the electric field gradient tensor [20]. Therefore,

$$\nu_{10} = \frac{Q_{10}}{Q_{11}} \frac{I_{11}(2I_{11}-1)}{I_{10}(2I_{10}-1)} \nu_{11} \tag{14}$$

and hence $\nu_{10} = 0.24$ MHz. Finally, the total nuclear quadrupole specific heat is the combination of $C_{10}$ and $C_{11}$, weighted according to their respective natural abundance, given by $C_Q = 6(x_{10}C_{10} + x_{11}C_{11})$. By using the nuclear quadrupole resonance frequencies from above, the nuclear quadrupole specific heat is found to be $C_Q = 2.30 \cdot 10^{-8}/T^2$ (J·mol$^{-1}$·K$^{-1}$), far too small to account for the observed upturn at low temperatures. At $T = 60$ mK, this would correspond to only 6.38 $\mu$J·mol$^{-1}$·K$^{-1}$, two orders of magnitude smaller than the measured value.

**Low-temperature thermal conductivity measurements.** Thermal conductivity of three SmB$_6$ crystals − two floating zone-grown, and one flux-grown − was measured at temperatures down to ≈ 150 mK and in magnetic fields up to 12 T (see figure in ref. [17]). A significant magnetic field enhancement in the low temperature thermal conductivity is seen especially in the floating zone-grown single crystals. The enhancement of the low temperature value of thermal conductivity in an applied magnetic field is a few orders of magnitude higher than the expectation associated with the electrical conductivity within a traditional Fermi liquid model, calculated using the Wiedemann-Franz relation and shown in the figure provided in ref. [17]. An increase in nuclear contribution to the specific heat capacity in a magnetic field would not be expected to contribute to the enhanced thermal conductivity, as it does not correspond to mobile excitations capable of carrying heat.

The enhancement of the low temperature thermal conductivity in finite magnetic fields in flux-grown crystals of SmB$_6$ is subtle compared to the magnetic field-induced enhancement in floating zone-grown crystals of SmB$_6$. A similarly low enhancement has also been reported in ref. [37], as shown in the figure provided in ref. [17]. Subtle differences in materials proper-ties between crystals prepared by the floating zone method and the flux growth technique are likely to be responsible for the observed difference below 1 K in thermal conductivity, the upturn at low temperatures in the quantum oscillation amplitude [13] and the linear specific heat coefficient (Fig. 2a). The smaller value of total thermal conductivity reported for a flux-grown crystals of $SmB_6$ in ref. [37] is consistent with the smaller sample thickness and hence a shorter mean free path compared to those of the samples measured here.

Very small disorder effects would also play a role in the suppression of the low temperature thermal conductivity. The high quality of our measured crystals is reflected in the large peak in high temperature thermal conductivity shown in the figure provided in ref. [17], which is considerably larger than those of previous generation samples [64]. Insulating materials exhibit a peak in the thermal conductivity where the phonon mean free path transitions from being limited by the sample boundaries at low temperatures, to being dominated by phonon-phonon scattering (Umklapp processes) at higher temperatures. The magnitude of this high temperature peak is strongly suppressed by lattice defects such as point defects, dislocations and stacking faults, and consequently it is a good indicator of sample quality [65].

Low temperature measurements of the thermal conductivity in this material are challenging because of the insulating character of this material, yielding large contact resistances. The large contact resistance between the sample and the thermal link results in a small temperature gradient across the sample. Measurements are hence very sensitive to factors such as thermometer calibration, particularly at low temperatures where the settling time for thermal equilibrium is rendered very long due to the high contact resistance. Another detrimental consequence of the high contact resistance is the tendency of phonons to thermally decouple due to the high contact resistance, as has been found for instance in the cuprate high temperature superconductors [66]. These effects impose a low temperature limit on the data, and we are careful with our measurements to only report results within the low temperature limit where such effects are minimised.

**Calculation of the thermal conductivity contribution from phonon transport.** The figure provided in ref. [17] shows the thermal conductivity of two floating zone-grown a single crystal of floating zone-grown SmB$_6$ in zero magnetic field, compared with the phonon contribution of the thermal conductivity calculated from kinetic theory, which relates the thermal conductivity $\kappa$ to the heat capacity $C_V$ via the equation

$$\kappa = \frac{1}{3} C_V d v_s \quad (15)$$

Here, $C_V$ is the heat capacity per unit unit volume, $d$ is the average sample dimension, and $v_s$ is the sound velocity of the material. The phonon contribution of the heat capacity at low temperatures is given by

$$C_V = \frac{12\pi^4}{5} \frac{k_B}{a_{\text{u.c.}}^3} \left(\frac{T}{\Theta_D}\right)^3 \quad (16)$$

We calculate the average sample dimension using $d = \sqrt{4tw/\pi}$, where $t$ is the thickness, and $w$ is the width of the sample. The sound velocity is given by

$$v_s = \frac{2k_B}{h} \Theta_D \left(\frac{\pi}{6n}\right)^{1/3} \quad (17)$$

where $h$ is the Planck constant, and $n$ is the number density of SmB$_6$, given by $n = a_{\text{u.c.}}^{-3}$. For a Debye temperature of $\Theta_D = 373$ K [55], we obtain a sound velocity of $v_s = 5179$ m/s. Expressing $\kappa/T$ as a function of $T^2$, we arrive at

$$\kappa/T = \alpha T^2 \quad (18)$$

where the gradient $\alpha$ is found to be $\alpha = 0.4772$ W·m$^{-1}$·K$^{-4}$ for the floating zone-grown crystal shown in Fig. 3b ($t = 0.43$ mm, $w = 0.23$ mm), and $\alpha = 0.5395$ W·m$^{-1}$·K$^{-4}$ for floating zone-grown crystal show in the inset of Fig. 3b ($t = 0.34$ mm, $w = 0.37$ mm). We find the total low temperature thermal conductivity in zero magnetic field to be described well by the calculated phonon contribution.

**Estimate of the effective mean free path.** Even for the best samples of SmB$_6$, quantum oscillations only become observable at significantly higher magnetic fields compared to metallic LaB$_6$, especially for the highest measured frequency, due to the much larger exponential damping term. The exponential damping term of the quantum oscillations, $R_D$, can be expressed in terms of the effective mean free path $l$ as

$$R_D = \exp\left(-\frac{B_0}{B}\right) = \exp\left(-\frac{\pi \hbar k_F}{eBl}\right) \tag{19}$$

where $B_0$ is given by the Dingle temperature, $T_D$, via $B_0 = \frac{2\pi^2 k_B m^*}{e\hbar} T_D$, and $k_F$ is the average Fermi wave vector, such that the effective mean free path is obtained as

$$l = \frac{\pi \hbar k_F}{eB_0} \tag{20}$$

At temperatures $\approx 1$ K and magnetic fields in the range $35 - 45$ T, we find that for the 11 kT frequency we have $k_F = \sqrt{2eF/\hbar} = 5.8 \cdot 10^9$ m$^{-1}$ and $B_0 \approx 200$ T for floating zone-grown samples. This gives a mean free path of $l \approx 5 \cdot 10^{-8}$ m in the magnetic field range $35$ T $\leq B \leq 45$ T for floating zone-grown samples. In the case of flux-grown samples, we find that the high frequency oscillations are significantly more suppressed in amplitude than for floating zone-grown samples due to a considerably higher exponential damping factor as revealed by their magnetic field dependence, making them much more challenging to observe.

To estimate the effective mean free path from the thermal conductivity, we use the formula presented in ref. [67], relating the thermal conductivity $\kappa$ to the scattering time $\tau$

$$\frac{\kappa}{T} = \frac{k_B^2 \tau}{m^* a_{\text{u.c.}}^3}, \tag{21}$$

where $a_{\text{u.c.}}$ is the lattice constant of SmB$_6$, $m^*$ is the effective mass in absolute units, and the scattering time is given by $\tau = \frac{l}{v_F}$, where $v_F$ is the Fermi velocity. We express $v_F$ in terms of the Fermi wave vector $k_F$ via $m^* v_F = \hbar k_F$. This results in an expression for the mean free path

$$l = \frac{\kappa}{T} \frac{\hbar k_F a_{\text{u.c.}}^3}{k_B^2}, \tag{22}$$

where $k_F = \sqrt{2eF/\hbar} = 5.8 \cdot 10^9$ m$^{-1}$ for the 11 kT frequency, and $a_{\text{u.c.}} = 0.413$ nm . At temperatures $\approx 0.2$ K and magnetic fields $\approx 12$ T, we find a value of $\kappa/T = 0.04$ W·m$^{-1}$·K$^{-2}$ for floating zone-grown samples from Fig. 3, giving a mean free path of $l \approx 9 \cdot 10^{-9}$ m in an applied magnetic field of 12 T. The lower value of the mean free path corresponding to thermal conductivity compared to the mean free path from quantum oscillations potentially reflects factors including a group velocity that is lower than the Fermi velocity due to a gapped charged sector, different itinerant length scales relevant to the two measurements, the effect of the lower magnetic fields at which the thermal conductivity measurements are performed, and the effect of thermal decoupling of phonons [66].

**Quantum oscillations in magnetisation.** Quantum oscillations in the magnetisation measured using capacitive Faraday magnetometry at the University of Tokyo are shown in the figure provided in ref. [17] in a field range from 7 to 14 T. Quantum oscillations in the magnetic susceptibility measured using extraction magnetometry in pulsed magnetic fields at the NHMFL Los Alamos are shown in the figure provided in ref. [17] in a field range from 29 to 65 T.

**Effective mass from quantum oscillations.** The effective mass of each of the Fermi surface orbits is obtained by mapping the temperature dependence down to 1 K, in which regime the temperature dependence is found to adhere to the Lifshitz-Kosevich form (see figure in ref. [17]). Below temperatures of $\approx 1$ K, an anomalous increase in quantum oscillation amplitude that displays a marked departure from Lifshitz-Kosevich form is observed in the majority of observed quantum oscillation frequencies in floating zone-grown samples (see figure in ref. [17]). A Lifshitz-Kosevich fit performed to the temperature dependence down to 1 K yields an effective mass which is in the range $0.1 \leq m^*/m_e \leq 1$ for the observed frequencies.

The figure provided in ref. [17] shows the derivative of the oscillatory magnetisation with respect to the temperature for the highest frequency which dominates the effective mass. The

Maxwell relation for the Helmholtz free energy is

$$V\left(\frac{\partial M}{\partial T}\right)_B = \left(\frac{\partial S}{\partial B}\right)_T \tag{23}$$

where $V$ is the volume of the crystal, $M$ is the magnetisation, and $S$ is the entropy. The finite value of the temperature derivative of the oscillatory magnetisation, and therefore of the entropy at low temperatures reveals the presence of the low-lying itinerant elementary excitations despite the charge gap in $SmB_6$.

**Theoretical models for quantum oscillations.** Encouragingly, the challenge to develop a complete theoretical model to capture the unconventional ground state of Kondo insulating $SmB_6$ as revealed by the entire suite of experimental results presented here has led to the exploration of new avenues including magnetic excitons, Majorana fermions, composite excitons, quantum oscillations arising from inside a filled band, quantum oscillations arising from open Fermi surfaces, an accompany-type valence fluctuation state, gapped charge quasiparticles and others [38, 39, 41, 43, 44, 68, 69, 70, 71, 72, 73, 74, 75, 76, 77, 78, 79].